\documentclass[english,10pt,aps,prd, a4paper,
preprintnumbers, 
floatfix,
nofootinbib,showpacs,superscriptaddress, notitlepage]{revtex4-1} 

\usepackage{graphicx}				
\usepackage{amssymb}
\usepackage{amsmath}
 \pdfoutput=1
\usepackage[usenames,dvipsnames]{color}  
\usepackage{setspace}
\usepackage{braket}
\usepackage{amsmath}
\usepackage{amssymb}
\usepackage[colorlinks=true,citecolor=darkred,urlcolor=darkred, pdfborder={0 0 0}]{hyperref}
\usepackage[normalem]{ulem}
\usepackage{xcolor}
\usepackage{amssymb,wasysym}

\makeatletter
\def\p@subsection{}
\makeatother

\definecolor{darkred}{rgb}{0.6,0,0}

\definecolor{linkcolor}{rgb}{0,0,0.5}

\usepackage{titlesec}
\titleformat*{\section}{\bfseries\boldmath\MakeUppercase}

\def\gsim{\raise0.3ex\hbox{$\;>$\kern-0.75em\raise-1.1ex\hbox{$\sim\;$}}}
\def\lsim{\raise0.3ex\hbox{$\;<$\kern-0.75em\raise-1.1ex\hbox{$\sim\;$}}}

\def\beqn#1{\begin{equation}\label{#1}}
\def\eeqn{\end{equation}}

\def\beqa#1{\begin{eqnarray}\label{#1}}
\def\eeqa{\end{eqnarray}}

\def\Z2{$\mathcal{Z_2}$}

\newcommand {\ignore}[1]{}

\def\321{$\mathrm{SU(3) \otimes SU(2) \otimes U(1)}$ }

\newcommand{\AddrUnina}{Dipartimento di Fisica E. Pancini, 
Universit\`a di Napoli Federico II \\
Complesso Universitario di Monte Sant'Angelo, Via Cinthia, Napoli (NA), Italy \\ and INFN, Sezione di Napoli.}
 \newcommand{\AddrUnisa}{Dipartimento di Matematica e Fisica, Universit\`{a} del Salento, and 
Istituto Nazionale di Fisica Nucleare, Sezione di Lecce, I-73100 Lecce, Italy.}

\begin{document}
  
\title{\color{BrickRed} Charged resonances and Minimal Dark Matter bound states at a multi-TeV muon collider}

\author{Natascia Vignaroli}\email{natascia.vignaroli@unisalento.it}
\affiliation{\AddrUnisa}\affiliation{\AddrUnina}

 \begin{abstract}
  \vspace{1cm} 
A multi-TeV muon collider proves to be very efficient not only for the search for new heavy neutral particles, but also for the discovery of charged bosons of the $W^\prime$ type. We find that, by analyzing the associated production with a Standard Model W, charged resonances can be probed directly up to multi-TeV mass values close to the collision energy, and for very small couplings with the SM fermions, of the order of $10^{-3}-10^{-4}$ times the SM weak coupling.
Additionally, charged bound states of WIMP Minimal Dark Matter, specifically a Majorana fermionic 5-plet, can be discovered with low statistics by running above the kinematic threshold, at a center-of-mass energy just slightly above the mass of the MDM bound state. This opens up a very interesting possibility for the discovery of WIMPs, complementary to the search for the resonant production of the neutral MDM bound state component, which relies on an on-peak search. For 5-plet MDM, indeed, the proposed search strategy is more efficient than the WIMP searches based on mono-X, missing-mass and disappearing tracks signatures.  

\end{abstract}
\maketitle

\newpage

{
  \hypersetup{linkcolor=BrickRed}
  \tableofcontents
}


\section{Introduction}

The design of a future $\mu^+ \mu^-$ collider (MuCol) with multi-TeV energy has recently been proposed, showing outstanding possibilities to discover and test different aspects of high energy physics \cite{MuonCollider:2022nsa,Black:2022cth,Accettura:2023ked}. Among these, very interesting prospects have been highlighted for the discovery of Weakly Interacting Massive particles (WIMPs) and for new heavy neutral bosons. 
\par For example, as shown in \cite{Huang:2021nkl}, $Z^\prime$ type of resonances from gauged $L_\mu - L_\tau$ models, produced through s-channel radiative return, can be probed directly up to very small couplings, of the order of $10^{-3}$, for a 1 TeV $Z^\prime$, by a 3 TeV MuCol with 1 ab$^{-1}$.
 A powerful test, and a higher reach on the mass of the resonance,  can also be obtained \textsf{indirectly}, via precision measurements. In this case, for example, a $Y$-universal $Z^\prime$ model can be ruled out by a 10 TeV MuCol, for $Z^\prime$ masses of the order of 100 TeV and couplings of the order of $10^{-1}$ \cite{Chen:2022msz}. 
\par WIMP dark matter can be realistically discovered by a multi-TeV MuCol up to the thermal target of the electroweak (wino-like) triplet, considering the channel where the WIMPs are produced in pairs, generating mono-X signals with large missing energy, or by the decays of heavier charged states of the WIMP EW multiplet, which give raise to disappearing tracks \cite{Capdevilla:2021fmj,Bottaro:2021snn}.
A powerful alternative strategy to test the WIMP scenario in its minimal realization, the minimal dark matter (MDM) hypothesis, 
is to consider the detection of bound states formed by two MDM fermionic weak multiplets \cite{Bottaro:2021srh}. For the Majorana 5-plet, 
MDM bound states can be produced resonantly with large cross section at a muon collider, provided the MuCol runs ``on-peak" at a center-of-mass energy  close to the mass of the bound state, $\sqrt{s}\approx 2M$. 
This would permit to discover MDM 5-plets with low statistics, few fb$^{-1}$, in the first phase of the collider operation \cite{Bottaro:2021srh}. Conversely, mono-X signals 
show lower sensitivities and would hardly be able to test the 5-plet MDM thermal target \cite{Han:2020uak}. A better coverage on WIMPs could be obtained by considering disappearing tracks \cite{Capdevilla:2021fmj} and missing-mass searches, even if the 5-plet target would be reached only for $\sqrt{s}\gtrsim 30\div 50$ TeV and a large amount of integrated luminosity, $L\gtrsim 2\div 100$ ab$^{-1}$ \cite{Bottaro:2021snn}. Precision measurements can also allow to test the 5-plet at a 14 TeV MuCol with $\mathcal{O}(10)$ ab$^{-1}$, even if only indirectly \cite{Franceschini:2022sxc}.
\par In this letter we propose a strategy that would allow to detect \textsf{directly} charged resonances of the $W^\prime$ type  at a muon collider, and a new efficient  way to test MDM bound states. 
This opportunity is offered by the analysis of the channel where the new charged state, either a $W^\prime$ or a MDM bound state, is produced in association with the $W$ boson of the Standard Model (SM). 
Besides the unique opportunity to directly test a $W^\prime$ new boson, with a very high reach, this analysis offers a very efficient way to test the WIMP scenario through the detection of MDM bound states, which is complementary to the search for the resonant production of the neutral bound state. 
Furthermore, it presents the advantage of not needing an ``on-peak" focus of the MuCol beam energy near the bound state mass, but just requiring the experiment to operate slightly above the kinematic threshold, $\sqrt{s}\gtrsim 2M$.   
\par In the following, after having introduced our theoretical framework for $X$ generic charged resonances of the $W^\prime$ type (Sec.~\ref{sec:w-prime}) and for the MDM bound states (Sec.~\ref{sec:MDM}), we will present our search strategy in Sec.~\ref{sec:search} and will offer our conclusions in Sec.~\ref{sec:conclusions}. 

\section{Charged resonances}\label{sec:w-prime}

We focus on a heavy spin-1 state transforming as a triplet under the SM electroweak group.
We consider the case of an effective $W^\prime$ boson, which we indicate as $X$, which interacts with the SM particles analogously to the SM $W$.
The relevant Lagrangian reads:
 \begin{equation}\label{eq:wp-eff}  
\mathcal{L}^{W^\prime}_{eff} = \frac{g_{X}}{\sqrt{2}}\left[ V^{CKM}_{ij}\bar u_i \gamma^\mu P_L d_j   +  V^{PMNS}_{ij}\bar \nu_i \gamma^\mu P_L \ell_j  \right] X_\mu + \text{H.c.} \, ,
\end{equation}
with $V^{CKM}$ and $V^{PMNS}$ denoting the CKM and PMNS matrices. In the case $g_{X}=g_2$, where $g_2=e/\sin\theta_w$ is the SM weak coupling, the $X$ interactions are identical to those of the SM $W$. This reproduces the Sequential SM scenario (SSM) of Ref. \cite{Altarelli:1989ff}. In our analysis we will leave $g_X$ as a free parameter. The $X$ decay rates read
\begin{align}\label{eq:decays}
\begin{split}
\Gamma(X^{\pm}\to \ell^{\pm}\nu)  \simeq\frac{g^2_X}{48 \pi}m_{X}  \quad
 \Gamma(X^{\pm}\to \bar{q}q')  \simeq \frac{g^2_X}{16 \pi}m_{X} \, .
\end{split}
\end{align}
The decays to Higgs and SM gauge bosons are suppressed.

The most stringent constraints on SSM $W^\prime$'s are currently set by the recent CMS search in the lepton plus missing transverse momentum final state \cite{CMS:2022krd}, at the 13 TeV LHC with 138 fb$^{-1}$, which excludes at 95\% C.L. a SSM $W^\prime$ lighter than 5.7 TeV. This limit reduces to softer bounds for $g_X<g_2$.

Charged  $W^\prime$'s also appear in composite Higgs theories \cite{Pappadopulo:2014qza, Vignaroli:2014bpa, Mohan:2015doa}, generated by a new strong dynamics as composite spin-1 weak triplet resonances. The composite $W^\prime$ couplings to Higgs and gauge bosons, differently from the SSM case, are typically relevant.

\section{Minimal Dark Matter Bound States}\label{sec:MDM}

A compelling and Minimal solution to the Dark Matter puzzle is to consider, in addition to the SM, a new fermionic multiplet under the SM weak gauge group \cite{Cirelli:2005uq}, whose neutral and, as such, lightest component constitutes a good DM candidate. The cosmological DM abundance can be then reproduced thermally, via freeze-out, for TeV-scale values of the DM mass, $M$ \cite{Cirelli:2005uq, Mitridate:2017izz}. For $M \gtrsim M_{W,Z}/\alpha_2$, pairs of MDM multiplets can form Coulombian-like electroweak bound states. This is verified for larger representations, and in particular in the case of 5-plets under $SU(2)_L$ of Majorana fermions with zero hypercharge. Note that the Majorana 5-plet represents a special case of MDM, because it can be made accidentally stable \cite{Cirelli:2005uq, Bottaro:2021snn}. 
After taking into account Sommerfeld and bound-state corrections \cite{Mitridate:2017izz}, the 5-plet thermal abundance matches the DM density for a mass $M \approx 14$ TeV.


Bound states with the same quantum numbers as the weak vectors inherit, via mixing, couplings to SM fermions. 
We are especially interested in such bound states, as they can
thereby be directly produced in $\mu^+ \mu^-$ collisions. In particular, the charged components of the bound state can be produced in the associated $W$ channel we are studying. 
For the 5-plet MDM, these special bound states exist: the spin-1 $^n s_3$ vector triplets \cite{Bottaro:2021srh}. The ground state $^1 s_3$, which will be the focus of our analysis, decays into fermions with a rate $\Gamma_{\text ann} = 15625\, \alpha_2^5 M/48 \approx 0.17$ GeV \cite{Bottaro:2021srh}. Its collider phenomenology
can be reproduced by the effective $W^\prime$ description of the previous section with an effective coupling:
\begin{equation}
g_X = g_{^1 s_3} \simeq 0.014 \, g_2 \, ,
\end{equation}
while its mass is $m_X=m_{^1 s_3}\approx 2 M \approx 28$ TeV.

\section{$\bf X^\pm W^\mp$ associated production }\label{sec:search}

\begin{figure}[]
\centering
\includegraphics[width=0.4\textwidth]{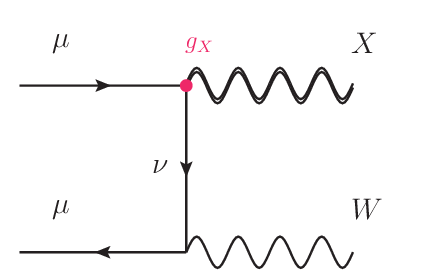}
\caption{\em Leading Feynman diagram for the SM $W$ associated production of a charged spin-1 resonance $X$ at a muon collider. }\label{fig:diag}
\end{figure}

We focus on the possibility to directly search for a charged spin-1 resonance $X$ at a muon collider, in the SM $W$ associated production channel $\mu^+ \mu^- \to X^\pm W^\mp$. The leading Feynman diagram is shown in Fig. \eqref{fig:diag}.

Note that in composite $W^\prime$ scenarios \cite{Pappadopulo:2014qza, Vignaroli:2014bpa, Mohan:2015doa} further contributions to the $XW$ associated production, which we will not include in our analysis, since they are suppressed for our SSM effective description, come from the exchange of an EW gauge boson in the $s$-channel. We will leave the analysis of these signal topologies to a future study. Therefore, the results we will show in this paper represent only a conservative estimate of the reach on a composite $W^\prime$ resonance.  

\begin{figure}[]
\centering
\includegraphics[width=0.4\textwidth]{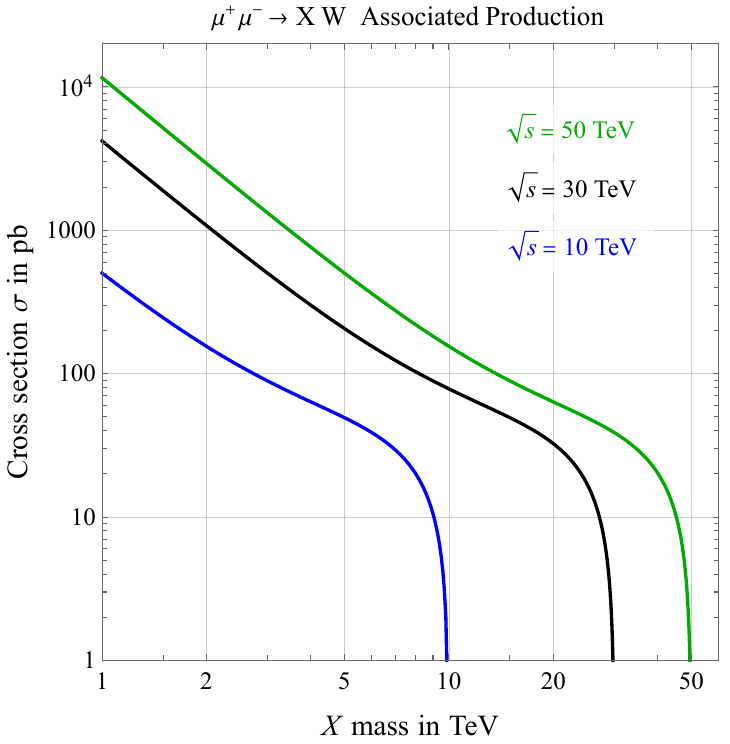} \qquad \includegraphics[width=0.4\textwidth]{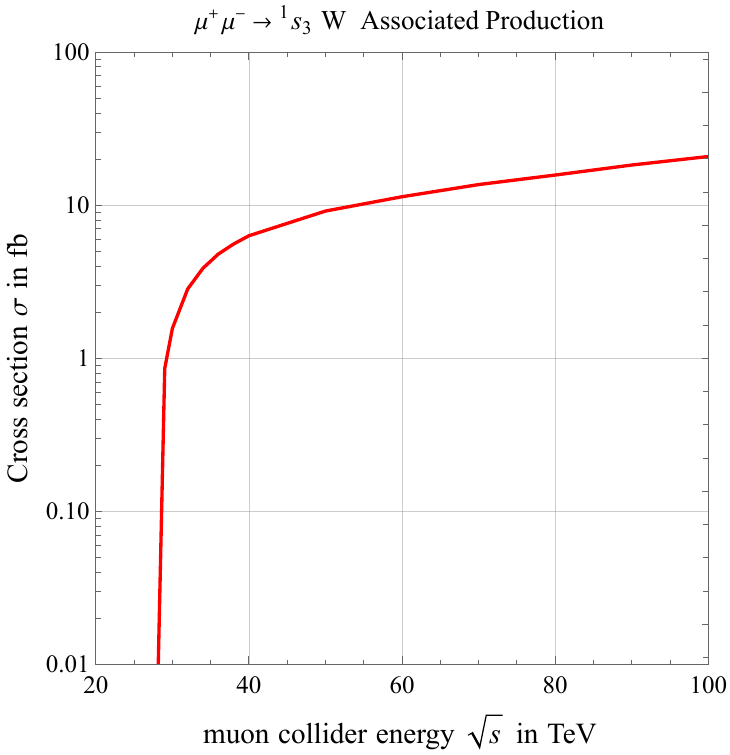}
\caption{\em Left plot: Cross section for the associated production of a charged spin-1 resonance $X$ with a SM $W$ (Fig. \ref{fig:diag}) at a muon collider with different beam energies, as function of the charged resonance mass. The plotted cross sections are calculated in the SSM case, with $g_X=g_2$. The cross section depends quadratically on $g_X$. Right plot: Cross section for the associated production of a 5-plet MDM bound state with a SM $W$ at a muon collider, as function of the center-of-mass energy.}\label{fig:xsec-wp}
\end{figure}

The cross section can be expressed in the analytic form:
\begin{align}
\begin{split}
&\sigma(\mu^+ \mu^- \to X^+ W^-) = \sigma(\mu^+ \mu^- \to X^- W^+) \simeq \\
&\frac{g^2_2 \, g^2_{X}}{1536\, \pi \, s^2 \, m^2_{X}\,  m^2_{W}} \big[  s^2 + 10 \, m^2_{X} s + m^4_{X} + m^4_{W}  \\ 
& + 10\, m^2_{W} (s- 5 m^2_{X})\big]\sqrt{(s-m^2_{X})^2-2 m^2_{W} (s+ m^2_{X})+ m^4_{W}}  \\
\end{split}
\end{align}
\noindent
We find good agreement between the analytic evaluation of the cross section and the numeric calculation with \texttt{MadGraph5} \cite{Alwall:2014hca}. The cross section is shown in Fig.~\ref{fig:xsec-wp} for different MuCol $\sqrt{s}$ (plot on the left), as a function of $m_X$, and, as a function of $\sqrt{s}$, in the specific case of the $W$ associated production of the $^1 s_3$ MDM bound state (plot on the right).\\
Note that EW radiation effects in high-energy lepton colliders may be significant for the very high energy cases above $\sqrt{s}=$10 TeV, as recently pointed out in \cite{Chen:2022msz} (see also \cite{Fadin:1999bq, Han:2020uid}). The evaluation of EW radiative corrections for the specific cases of the $W^\prime$ and MDM bound state considered in this manuscript deserves a dedicated investigation, which we leave for future studies.~\footnote{However, naively, one can expect that, since the charged resonances considered in this study are of the SSM type and have only suppressed couplings to the SM gauge bosons, corrections to the EW leading-order calculations presented in the manuscript might be be less significant than those reported in \cite{Chen:2022msz} for the $WW$ production process.} 
Despite non negligible corrections might be expected in the cases of very high collision energies, we think it is useful to report our results based on EW leading-order calculations also for the case of $\sqrt{s}=50$ TeV.

\subsection{Selection strategy and reach}

We consider the fully hadronic final state $\mu^+ \mu^- \to (W \to j j) (X \to j j)$. 
The SM $W$ decay products, as a consequence of the large Lorentz boost, are emitted very collimated, and will be mostly collected in a single jet. \footnote{The fraction of resolved jets from the $W$ (with $\Delta R>$ 0.4) becomes non negligible, reaching values of the order of 18\%, only for the heaviest $X$ cases, near the kinematic limit $m_X \approx \sqrt{s}$. For these limit values, a slight modification to the signal selection strategy could be applied in order to include these events and improve the sensitivity.} 

We thus require at least 3 hard jets in the central region, with sufficient separation from each other:
\begin{equation}\label{eq:accept}
p_T\, j > 30 \;  \text{GeV} \, , \quad  |\eta_j|< 2.5\, , \quad  \Delta R_{jj} > 0.4 \; ,
\end{equation}
where $\Delta R_{jj} = \sqrt{\Delta \phi_{jj} ^2 + \Delta \eta_{jj}^2}$ denotes the angular separation between two jets. We also assume a detection efficiency of 70$\%$ for each jet in this acceptance region. \par
Signal and background events are simulated with \texttt{MadGraph5} \cite{Alwall:2014hca}. Events are then passed to \texttt{Pythia8} \cite{Bierlich:2022pfr} for showering. Jets are clustered with \texttt{Fastjet} \cite{Cacciari:2011ma} by using an anti-kt algorithm with cone size $R=0.4$. We also apply a smearing to the jet 4-momenta, following the \texttt{Delphes} \cite{deFavereau:2013fsa} default card, in order to minimally take into account detector effects. \footnote{According to studies on the expected performance of a future muon collider, even better detector performances, than those considered in this analysis, can be envisaged \cite{MuonCollider:2022ded}. However, we prefer to remain conservative in our predictions.} 
\par
The SM background, which we find to be of the order of a few fb in the acceptance region, is given mainly by events with jets emitted by the radiation of an s-channel photon, where the third jet is produced by a gluon radiated from a quark. A subdominant background component consists of jets from $WW$ production, while the background contribution from $t \bar t$ events is almost negligible.
\par
\begin{figure}[]
\centering
\includegraphics[width=0.32\textwidth]{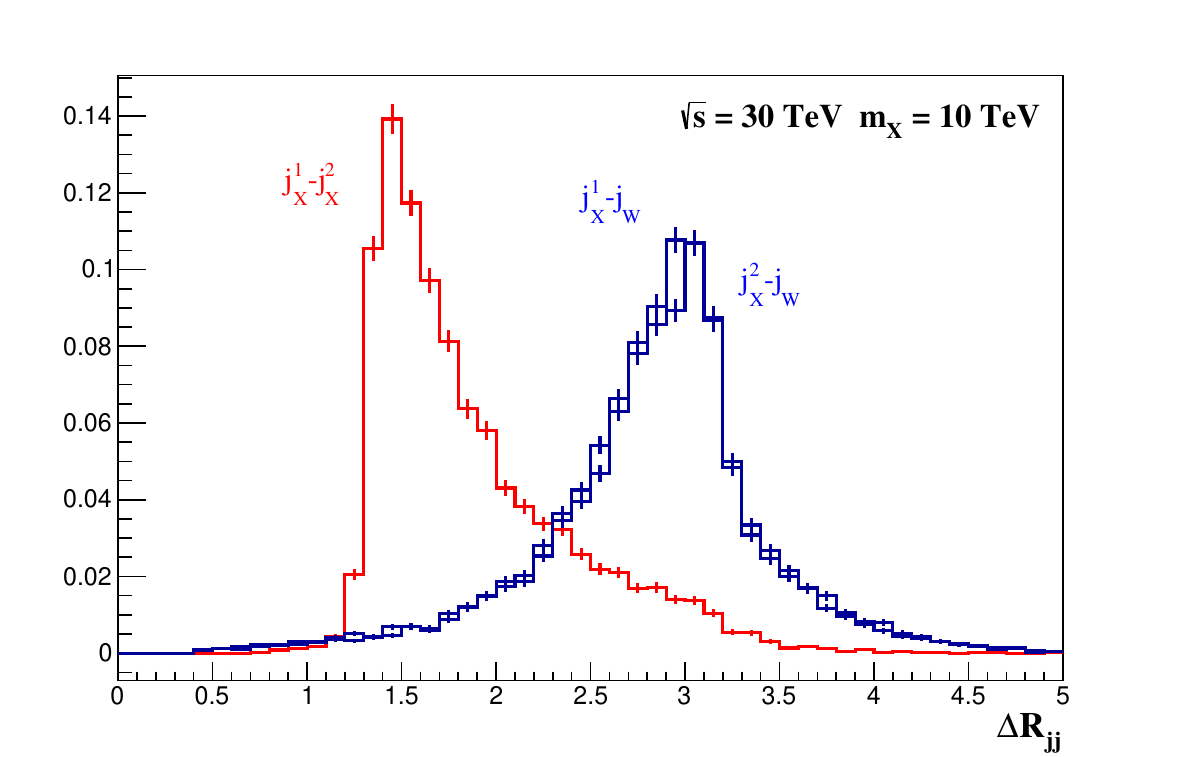}
\includegraphics[width=0.32\textwidth]{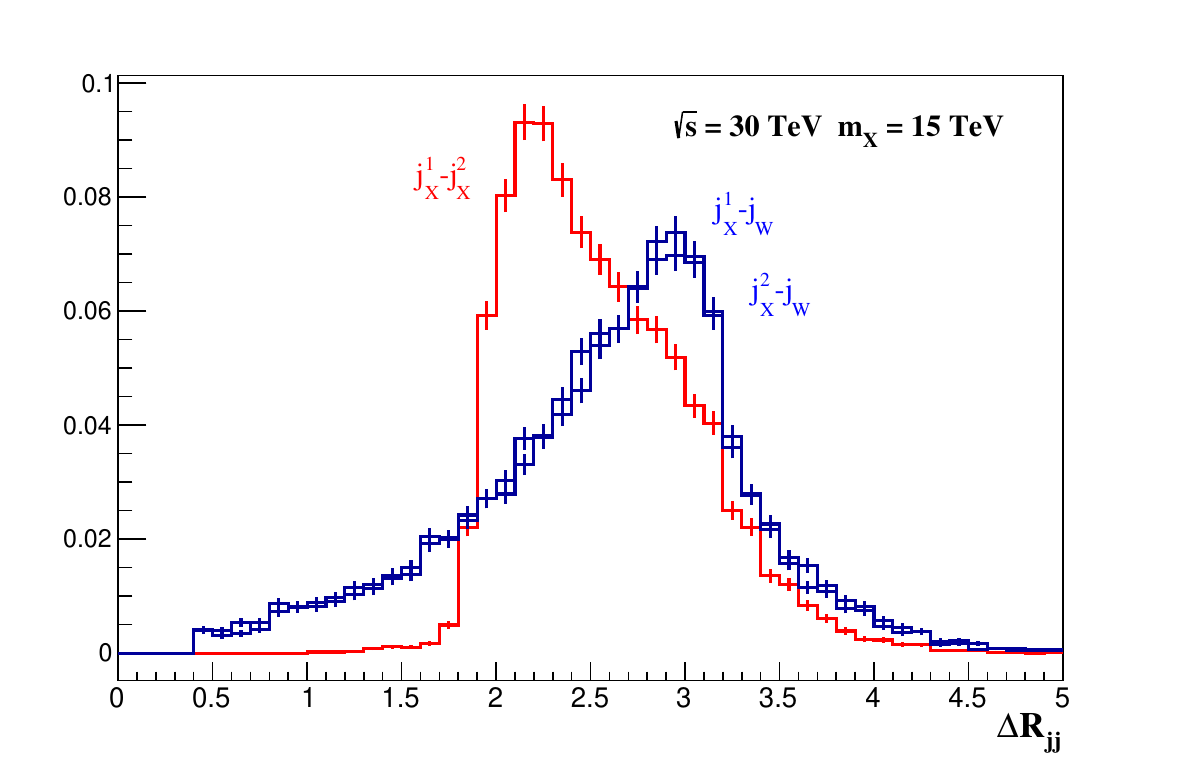}
\includegraphics[width=0.32\textwidth]{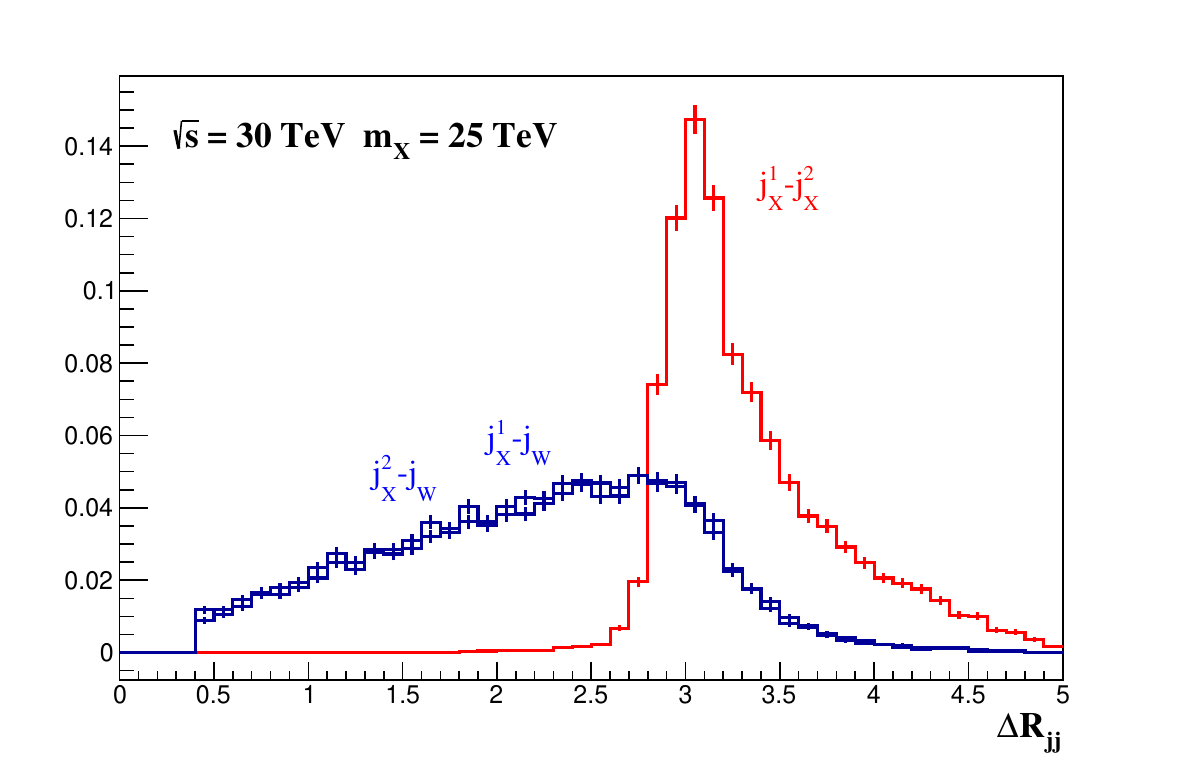}
\caption{\em From Monte Carlo truth: Signal event distributions (normalized to unit area) in the $\Delta R$ separation between two jets: between the two jets coming from the $X$ decays (in red), between the single jet from the $W$ decay and one of the two jets from the $X$ decay (the two histograms in blue). We consider $\sqrt{s}=30$ TeV. The plot on the left refers to a lighter $X$ scenario, with $m_X=10$ TeV, the plot on the right to a heavier case with $m_X=25$ TeV.  The middle plot refers to an intermediate case with  $m_X=\sqrt{s}/2=15$ TeV.  }\label{fig:deltaR}
\end{figure}

We apply a simple strategy to identify the jets coming from the $X$ decays, and consequently to reconstruct the heavy charged resonance. We observe (cfr. Fig. \ref{fig:deltaR}) that in the case of heavier $X$ resonances, the jets from the $X$ are mostly emitted back-to-back and have a large $\Delta R$ separation; for lighter $X$, instead, they tend to be collimated and closer. 
As shown in Fig. \ref{fig:deltaR}, in the case of heavy $X$ resonances, $m_X>\sqrt{s}/2$, the two jets emitted by the $X$ decays have a large $\Delta R$ separation. For the majority of the events, this separation is the largest one among the $\Delta R$'s of the three $p_T$-leading jets. For lighter $X$, instead, the two emitted jets are collimated and closer. In this case, for most of the events, their separation is the smallest one. 
We thus consider the three $\Delta R$ separations among the three $p_T$-leading jets and, for $m_X \leq \sqrt s/2$, we identify the two jets from the $X$ by assuming that they are those with the smallest separation (consequently the third remaining jet is identified with the $W$ jet), while, for $m_X > \sqrt s/2$, the two $X$ jets are identified with those with the largest separation (the third remaining jet constitutes the hadronically decayed $W$).

In order to reduce the background and to obtain a clean reconstruction of the $X$ resonance, we then consider a cut on the invariant mass of the reconstructed $W$ boson ($M_W$): 
\begin{equation}\label{eq:mw-cut}
50\, \text{GeV}< M_W < 110\, \text{GeV} \, .
\end{equation}
The efficiencies of this $W$ and $X$ reconstruction strategy range from 98\% for the lightest $X$ values to about 40\% for $m_X \simeq \sqrt s/2$ in the lower $X$ mass case, and from about 67\%  to 90\% in the case of heavier $X$. Instead, the background is substantially reduced to negligible levels (cfr.~Table \ref{tab:cut-flow}, which shows the cut flow of the cross sections). The drop in sensitivity for $m_X\simeq\sqrt{s}/2$ (which will reflect in a slight deficit on the final reach for this region) is due to the lower reconstruction efficiency of the $X$ and $W$ resonances for this specific kinematic configuration, as evident from the central plot in Fig. \ref{fig:deltaR}: for $m_X\simeq\sqrt{s}/2$, the $\Delta R$ separations between the two jets from the $X$ and between one of the $X$ jet and the single jet from the $W$ tend to overlap, so that their distinction becomes less efficient. 
The sensitivity in this specific region can be improved by exploiting different reconstruction strategies. The reconstruction strategy based on the $\Delta R$ separation we apply is conservative. More refined techniques, indeed, could be also applied, which we leave for future investigation. For example, one could identify the different jets origin by analyzing the  invariant mass and structures of the final state jets or of their combinations. We prefer, however, to rely on the simple strategy described above, since it shows already good efficiencies and, moreover, it is much less dependent on still-unknown detector performance details and on yet-to-be-tuned modeling of jet showering effects.

\begin{table}[]
\centering
{\footnotesize
\begin{tabular}{|c|c|c|}
\hline 
 & &   \\
 & &   \\[-0.6cm]
 \textsf{$\sqrt{s}=$ 10 TeV} & accept. & $W$ reco.  \\  [0.2cm]
 \hline
 & &   \\[-0.1cm]
  $m_X$ (TeV) & &  \\ [0.25cm]
 5  & 4.66 & 1.94        \\[0.25cm]
 6  & 3.63 &  1.19      \\[0.25cm]
 7  & 2.69 &  2.14       \\[0.25cm]
 8  & 1.82 &  1.67       \\[0.25cm]
 9  & 0.939 &  0.897       \\[0.25cm]
 9.9  & 0.177  &  0.152       \\[0.25cm]
  &   &         \\[0.25cm]
   &   &         \\[0.25cm]
\hline
 & &  \\
 & &   \\[-0.6cm]
  $Z/\gamma^* \to jets$            & 1.95 $\cdot 10^{-3}$ & 1.49 [1.26] $\cdot 10^{-4}$  \\[0.17cm]
  $VV \to jets$        & 0.077 $\cdot 10^{-3}$ & 0.27 [0.58] $\cdot 10^{-4}$   \\[0.17cm]
  Total                  & &  \\
  background      &  2.03 $\cdot 10^{-3}$ &  1.76 [1.84] $\cdot 10^{-4}$    \\[0.15cm]
\hline
\end{tabular}
}
{\footnotesize
\begin{tabular}{|c|c|c|}
\hline 
 & &   \\
 & &   \\[-0.6cm]
 \textsf{$\sqrt{s}=$ 30 TeV} & accept. & $W$ reco.  \\  [0.2cm]
 \hline
  & &   \\[-0.1cm]
  $m_X$ (TeV) & &  \\ [0.25cm]
 5 & 20.3 & 19.9         \\[0.25cm]
 10 & 7.53 & 6.23      \\[0.25cm]
 15  & 4.65 & 1.93        \\[0.25cm]
 20   & 2.97 &  1.97       \\[0.25cm]
 25  & 1.52 &  1.40       \\[0.25cm]
 28  & 0.640 &  0.603       \\[0.25cm]
 29.9  & 0.151  &  0.137       \\[0.25cm]
  &   &         \\[0.25cm]
\hline
 & &  \\
 & &   \\[-0.6cm]
  $Z/\gamma^* \to jets$            & 0.246 $\cdot 10^{-3}$ & 1.84 [1.17] $\cdot 10^{-5}$  \\[0.17cm]
  $VV \to jets$        & 3.2 $\cdot 10^{-7}$ & 0.68 [1.6] $\cdot 10^{-7}$   \\[0.17cm]
  Total                  & &  \\
  background      &  0.246 $\cdot 10^{-3}$ &  1.84 [1.18] $\cdot 10^{-5}$    \\[0.15cm]
\hline
\end{tabular}
} \\[0.2cm]
{\footnotesize
\begin{tabular}{|c|c|c|}
\hline 
 & &   \\
 & &   \\[-0.6cm]
 \textsf{$\sqrt{s}=$ 50 TeV} & accept. & $W$ reco.  \\  [0.2cm]
 \hline
  & &   \\[-0.1cm]
  $m_X$ (TeV) & &  \\ [0.25cm]
 5 & 49.8 & 49.6         \\[0.25cm]
 10 & 15.2 & 14.7      \\[0.25cm]
 20  & 6.11 & 4.21        \\[0.25cm]
 30  & 3.62 &  1.15       \\[0.25cm]
 35  & 2.70 &  2.12       \\[0.25cm]
 40  & 1.80  &  1.64       \\[0.25cm]
 45  & 0.937  &  0.874      \\[0.25cm]
 49.9  & 0.146  &  0.132       \\[0.25cm]
\hline
 & &  \\
 & &   \\[-0.6cm]
 $Z/\gamma^* \to jets$            
 & 9.10 $\cdot 10^{-5}$ & 6.60 [3.82] $\cdot 10^{-6}$  \\[0.17cm]
  $VV \to jets$        
 & $<$ 1 $\cdot 10^{-7}$ & $<$ 1 $\cdot 10^{-8}$  \\[0.17cm]
  Total                 
 & &  \\
 background      
&  9.10 $\cdot 10^{-5}$ & 6.60 [3.82] $\cdot 10^{-6}$    \\[0.15cm]
\hline
\end{tabular}
}
\caption{\label{tab:cut-flow} \em Cross section values (in pb) for signal (assuming $g_X=g_2$) and background, in the acceptance region, and after the reconstruction of the $W$ boson, including the cut in Eq.~\eqref{eq:mw-cut}. $V$ denotes a SM weak gauge boson, $V\equiv Z, W, \gamma^*$. For the background, the values on the third column refer to those following the reconstruction strategy for heavier [lighter] $X$ resonances.
}
\end{table}

\begin{figure}[]
\centering
\includegraphics[width=0.47\textwidth]{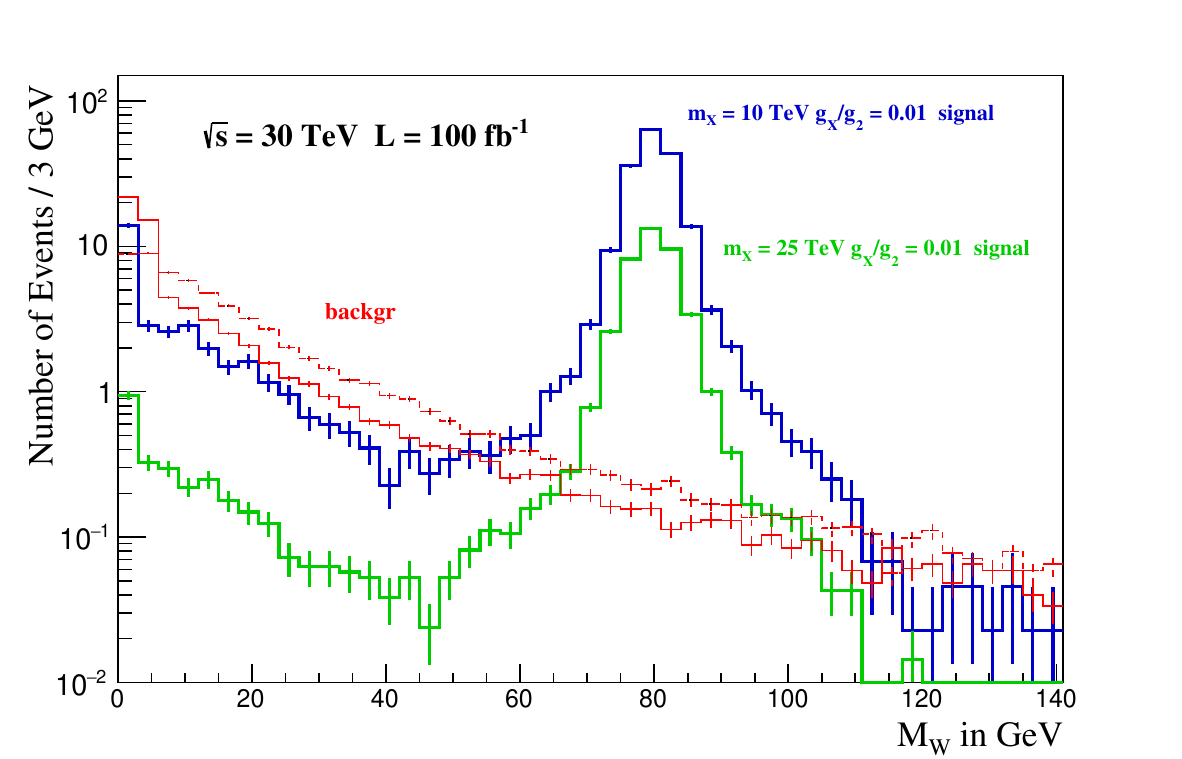}
\includegraphics[width=0.47\textwidth]{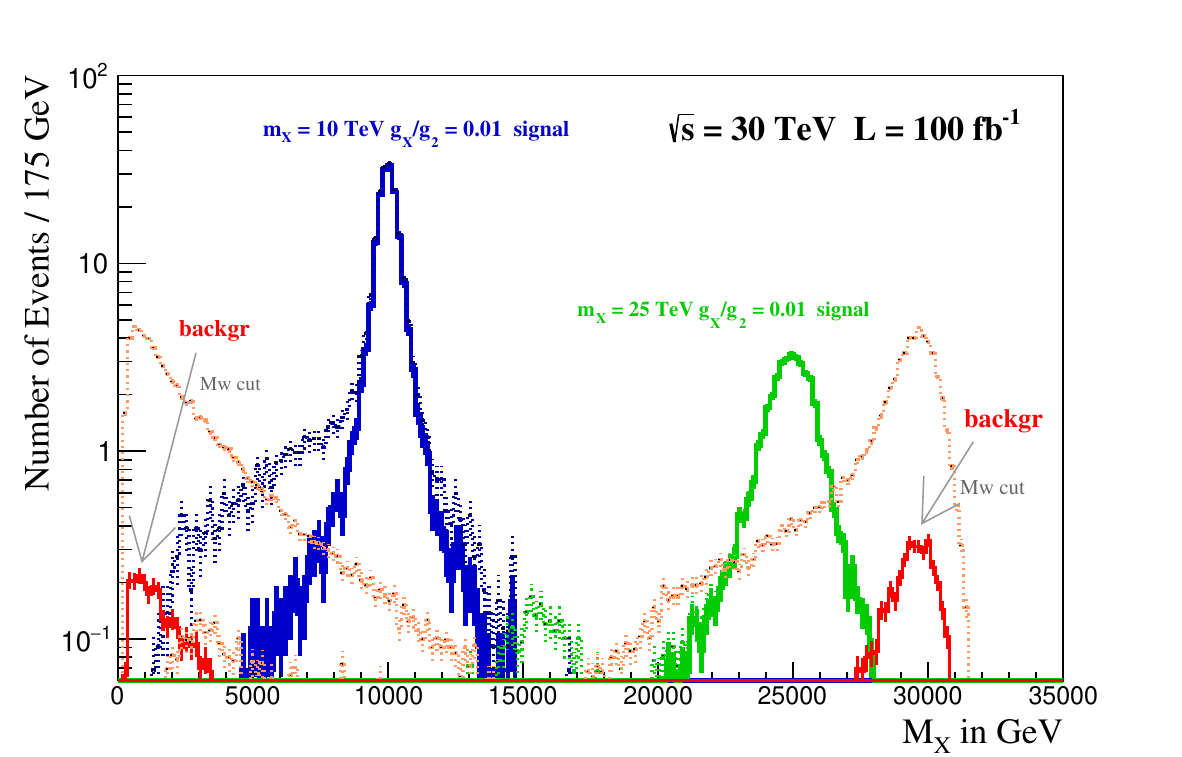}
\caption{\em Event distributions, at a 30 TeV muon collider with an integrated luminosity of 100 fb$^{-1}$, for the signals of an effective $W^\prime$ of 10 and of 25 TeV, with a coupling $g_X=0.01 g_2$, and for the background, resulting from the two cases of a lighter and a heavier $X$ selection strategy. Plot on the left: Invariant mass distribution of the reconstructed $W$ boson. Plot on the right: Invariant mass distribution of the reconstructed $X$ resonance. The signal and background distributions 
are shown before (in dashed lines) and after (continuous thick lines) the cut on the invariant mass of the reconstructed $W$ boson, Eq. \eqref{eq:mw-cut}.}\label{fig:InvMass}
\end{figure}

\par
\begin{figure}[]
\centering
\includegraphics[width=0.5\textwidth]{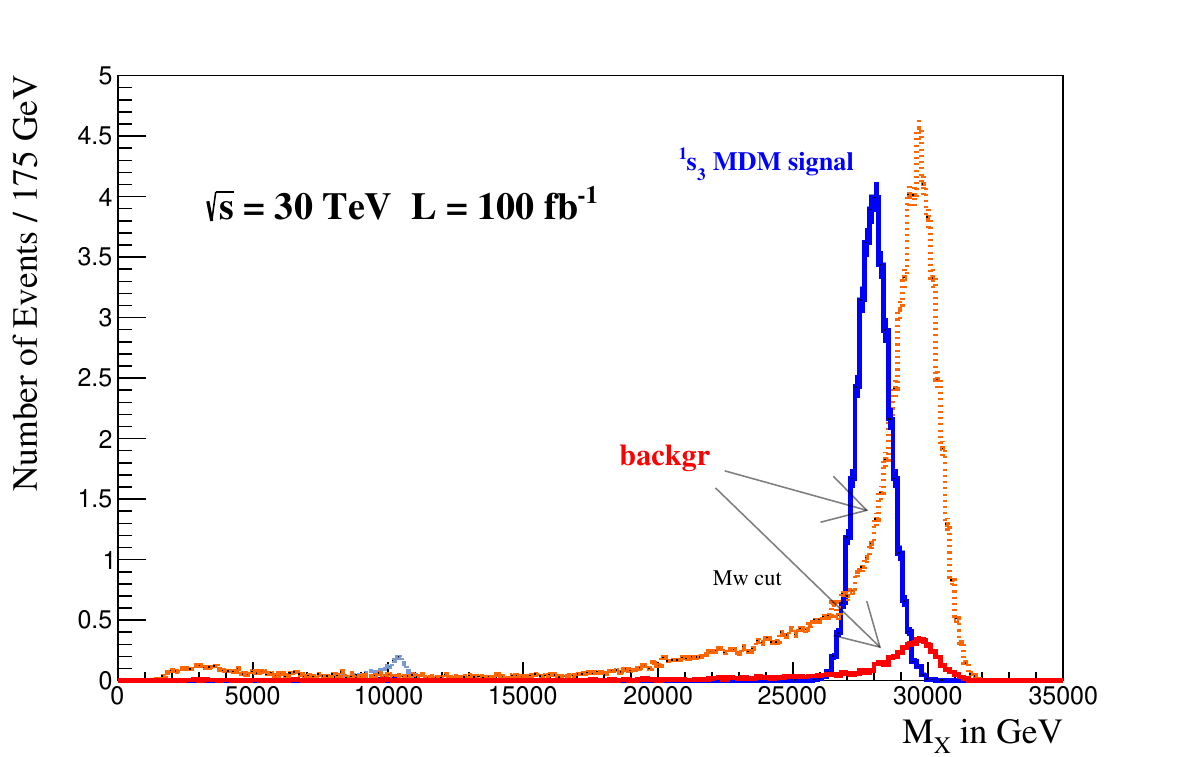}
\caption{\em Invariant mass distribution of the reconstructed $X$ resonance for the signal of a $^1 s_3$ MDM bound state and for the background, before (in dashed lines) and after (continuous thick lines) the cut on the invariant mass of the reconstructed $W$ boson, Eq. \eqref{eq:mw-cut}. The plot shows the event distribution at a 30 TeV muon collider with an integrated luminosity of 100 fb$^{-1}$. }\label{fig:X-reso}
\end{figure}

Fig. \ref{fig:InvMass} shows the $W$ and the $X$ invariant mass distributions, at a 30 TeV muon collider, for the signals of an effective $W^\prime$ resonance in two cases corresponding to a lighter ($m_X=10$ TeV) and a heavier ($m_X=25$ TeV) $X$ scenario, and for the background. The $M_X$ distributions are shown before, in dashed lines, and after, in continuous thick lines, the cut on the reconstructed $W$ invariant mass, Eq.~\eqref{eq:mw-cut}.  
The analysis described above can be applied as well to the case of the charged component of the MDM bound state $^1 s_3$. The corresponding $M_X$ invariant mass distribution, at a 30 TeV muon collider, is shown in Fig. \ref{fig:X-reso} together with the background distribution, before and after the cut on $M_W$. \footnote{Note that the small peak in dashed blue line around 10 TeV in Fig. \ref{fig:X-reso}, as also the peak in dashed green in the right plot of Fig. \ref{fig:InvMass} around 15 TeV, are generated by a small fraction of signal events for which the $X$ and $W$ reconstruction fails. More specifically, for a small number of signal events, the $W$ is non-correctly reconstructed and it is found to have a mass typically smaller than the real $W$ mass; for the same events, the $X$ is non-correctly reconstructed as well, generating a peak at a lower mass. This small peak disappears once the cut on the $W$ invariant mass is applied. }  
It is evident from Fig.s \ref{fig:InvMass} and \ref{fig:X-reso} how, after the complete signal selection strategy, the $^1 s_3$ and the effective $W^\prime$ resonances are clearly distinguishable from the background.

\par
\begin{figure}[h!]
\centering
\includegraphics[width=0.47\textwidth]{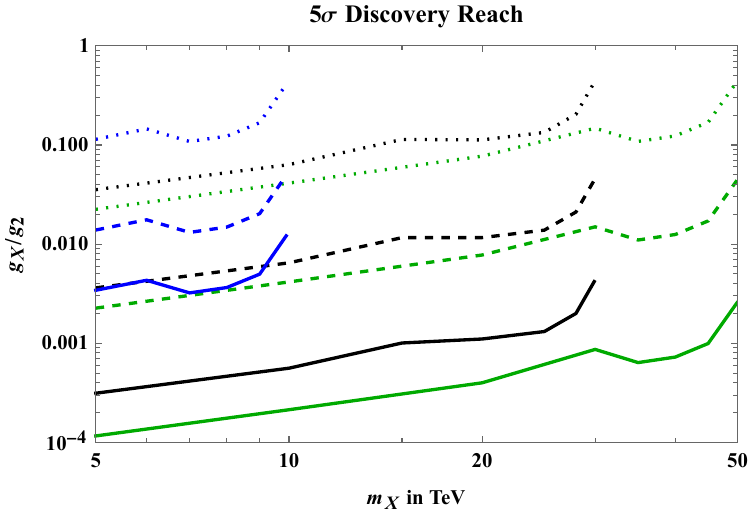}\,
\includegraphics[width=0.47\textwidth]{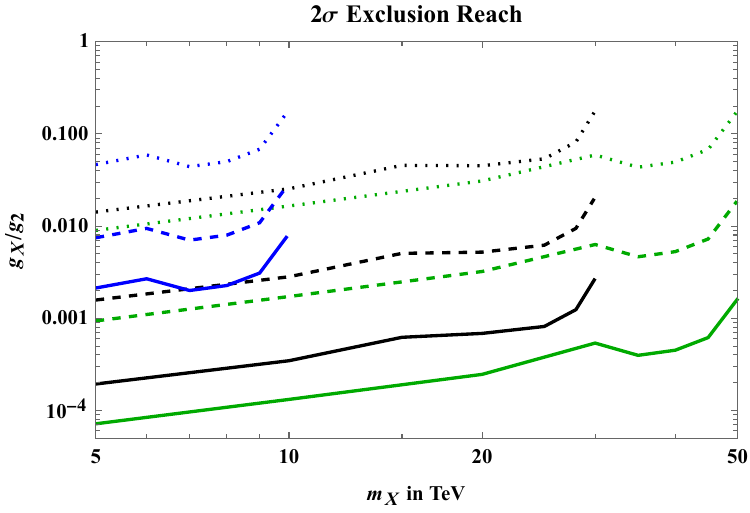}
\caption{\label{fig:reach} \em 5$\sigma$ discovery (left plot) and 2$\sigma$ exclusion (right plot) reach on the plane mass of the $X$ versus coupling, expressed in terms of the SM weak coupling, for a 10 TeV (blue curve), 30 TeV (black) and 50 TeV (green) MuCol with a collected integrated luminosity of 1 fb$^{-1}$ (dotted curves), 100 fb$^{-1}$ (dashed) and for the maximum achievable luminosity (continuous curves).}
\end{figure}

The final reach of our analysis on a generic $X$ charged resonance is shown in Fig.~\ref{fig:reach}.  We have estimated the statistical significance as $S/\sqrt{S+B}$, with $S$ ($B$) denoting the signal (background) number of events.   We indicate the 5$\sigma$ discovery and the 2$\sigma$ exclusion reach for a 10, 30 and a 50 TeV muon collider with different collected integrated luminosities, up to the maximum achievable value, $L=10\, (\frac{\sqrt{s}}{10\, \text{TeV}})^2$ ab$^{-1}$ \cite{Han:2020uak}. \footnote{The reported reach does not include systematic errors. By including conservatively a systematic uncertainty of 10\% \cite{MuonCollider:2022ded} on both the signal and the background, we estimate that the sensitivities to the $g_X$ coupling decrease by no more than 5\% compared to the values shown in Fig.~\ref{fig:reach}. Note that, in general, the search channel and the strategy considered in this study, enjoying a very high signal-to-background ratio, are little affected by systematics.} The MuCol can probe with this analysis in the associated $WX$ channel, charged $X$ resonances up to mass values close to the center-of-mass energy, and for couplings as small as $10^{-2} g_2$, $10^{-3} g_2$ and $10^{-3}-10^{-4} g_2$ for $\sqrt{s}=$10, 30 and 50 TeV respectively. Furthermore, a $W^\prime$ in the SSM scenario, corresponding to the case $g_X/g_2=1$, can be discovered at the very early stage of running by a muon collider. We find that \textsf{with just 50 pb$^{-1}$ of integrated luminosity, a SSM $W^\prime$ with a mass up to 9, 28 and 46 TeV can be discovered by the MuCol with $\sqrt{s}=$ 10, 30 and 50 TeV respectively.} 
This marks an unprecedented level for a direct search. For comparison, the efficiency that can be achieved in a direct search for a heavy resonance at the future 100 TeV proton-proton collider, FCC-hh \cite{FCC:2018vvp}, is about one-to-two orders of magnitude lower than the efficiency of the proposed search at a 10 TeV muon collider (cfr. \cite{Golling:2016gvc, Thamm:2015zwa}).\\
In the case of the \textsf{MDM bound state}, we find the following expected reach for a 5-plet MDM $^1 s_3$ charged bound state of 28 TeV: \textsf{ it can be excluded with about 34 fb$^{-1}$ and discovered with 210 fb$^{-1}$ by a 30 TeV muon collider}. This muon collider reach is even higher for larger collision energies and/or mass values lower than 28 TeV. That is, for $\sqrt{s}/m_{^1 s_3}\gtrsim 1.07 $, the values we indicate represents a conservative estimate of the MuCol reach on a 5-plet MDM bound state of mass $m_{^1 s_3}$. 
The reach found in this paper on the 5-plet MDM is significantly more efficient than the reach of WIMP searches based on missing-mass and disappearing tracks signatures, which would be able to test the 5-plet target only for $\sqrt{s}\gtrsim 30\div 50$ TeV and a large amount of integrated luminosity, $L\gtrsim 2\div 100$ ab$^{-1}$ \cite{Capdevilla:2021fmj, Bottaro:2021snn}. Even lower sensitivities are expected from mono-X searches, which would need very high collision energies, around 100 TeV, in order to reach the 5-plet target \cite{Han:2020uak}.

\section{Conclusions}\label{sec:conclusions}

In this letter we have proposed a new channel and strategy to probe \textsf{directly} heavy charged resonances at a future multi-TeV muon collider: The associated production of the charged new state with a SM W.
The projected sensitivities of the MuCol in the channel are shown in Fig.~\ref{fig:reach} and indicate that a  
charged resonance of the SSM $W^\prime$ type\footnote{As explained, the reach of this analysis is expected to be conservative in the case of $W^\prime$ resonances from composite Higgs theories.} can be discovered up to multi-TeV mass values close to the beam-colliding energy, and for very small couplings with the SM fermions, of the order of $10^{3}-10^{ 4}$ times smaller than the SM weak coupling. This sensitivity level would be unprecedented for a direct search.
Furthermore, the channel offers a very efficient and alternative way to probe the WIMP scenario for the special case of MDM in the 5-plet EW representation, by allowing the direct detection of the charged component of the MDM bound state. A MDM Majorana 5-plet bound state can be excluded with about 34 fb$^{-1}$ and discovered with 210 fb$^{-1}$ by a 30 TeV MuCol.
This reach on the WIMP 5-plet thermal target is much higher than those of mono-X, missing-mass and disappearing tracks signatures.

\acknowledgments

The author thanks Salvatore Bottaro and Alessandro Strumia for previous collaboration on related topics, and Roberto Franceschini  for discussions leading to this work and comments on the manuscript. This work was partially supported by ICSC – Centro Nazionale di Ricerca in High Performance Computing, Big Data and Quantum Computing, funded by European Union – NextGenerationEU, reference code CN\_00000013.

\bibliography{biblio.bib}

\end{document}